\documentclass[%
 reprint,
superscriptaddress,
showpacs,
preprintnumbers,
nobibnotes,
 amsmath,amssymb,
 aps,
prl
]{revtex4-1}
\usepackage{url} %this package should fix any errors with URLs in refs.
\usepackage{amssymb}
\usepackage{amsbsy}
\usepackage{amsmath}
\usepackage{color}
\usepackage{psfrag,epsfig}
\usepackage{xcolor,graphicx}
\graphicspath{{./figures/}{./}}
\usepackage{times}
\usepackage{mathptmx}

\begin{document}

%\preprint{APS/123-QED}
\title{Three-dimensional Vortex-induced Reaction Hotspots at Flow Intersections}% Force line breaks with \\

\author{Sanghyun Lee}
\affiliation{Department of Earth and Environmental Sciences, University of Minnesota, Minneapolis, USA}
\author{Peter K. Kang}%
\email[Corresponding author: ]{pkkang@umn.edu}
\affiliation{Department of Earth and Environmental Sciences, University of Minnesota, Minneapolis, USA}
\affiliation{Saint Anthony Falls Laboratory, University of Minnesota, Minneapolis, USA}

\date{\today}% It is always \today, today,
%Since 2011 Letters are no longer limited to four pages but rather to a maximum word count, currently 3750 words, not including abstract.
\begin{abstract}
We show the emergence of reaction hotspots induced by three-dimensional (3D) vortices with a simple $A+B \rightarrow C$ reaction. We conduct microfluidics experiments to visualize the spatial map of the reaction rate with the chemiluminescence reaction and cross-validate the results with direct numerical simulations. 3D vortices form at spiral saddle type stagnation points, and the 3D vortex flow topology is essential for initiating reaction hotspots. The effect of vortices on mixing and reaction becomes more vigorous for rough-walled channels, and our findings are valid over wide ranges of channel dimensions and Damk\"{o}hler numbers.
\end{abstract}
% Secondary publications and information retrieval purposes.
%\pacs{47.56.+r, 47.60.+i, 47.15.−x, 05.10.Gg, 05.40.Fb, 05.60.−k}
% You may use the \texttt{description} environment to structure your abstract;
% use the optional argument of the \verb+\item+ command to give the category of each item. 

\maketitle

% Intro
% Importance of vortex
Vortices commonly occur in various channel flow systems such as rock fractures~\citep{kosakowski1999, boutt2006trapping, cardenas2009effects,lee2015tail}, porous media~\citep{Wood2007,ye2015experimental,crevacore2016recirculation,pasquier2017}, pipe flows~\citep{Ault2016,oettinger2018}, micromixers~\citep{Xi7516}, and blood vessels~\citep{Gharib2006,Biasetti2011}. Specifically, vortices can have a distinctive flow topology~\citep{delery2001,haller2001,Bresciani2019}, and the topology of a flow field is known to control mixing processes, which in turn control reaction dynamics~\citep{de2012flow,engdahl2014predicting,turuban2018space}. Vortices at fluid flow intersections are particularly important because fluids with different properties can mix and react at flow intersections~\citep{lee2016passive,Zou2017}. Notably, vortices may alter mixing dynamics and initiate local reaction hotspots where reaction rates are locally maximum. Nevertheless, to the best of our knowledge, there has been no study that elucidated the role of three-dimensional (3D) vortices on mixing and reaction at flow intersections.

%Recently, a study revealed that the connected advective flow paths between vortices and main flow channel, nature of the 3D vortex flow topology, can exert dominant control over mass transfer processes~\citep{lee2015tail}. As such, vortices may alter mixing dynamics and, consequently, initiate local reaction hotspots where reaction rates are locally maximum. However, numerous present-day studies were limited to two-dimensional (2D) systems~\citep{zhou2019mass,Dou2018} and to the best of our knowledge, there has been no study that elucidated the role of 3D vortices on mixing and reaction at flow intersections. Furthermore, the impacts of such localized phenomena should be taken into account in generating an accurate representation of the entire system~\citep{Battiato2011}.
%Flow topology of 2D vortices and 3D vortices are fundamentally different~\citep{kang2019}, 

In this study, we combined laboratory microfluidic experiments and direct numerical simulations to establish a previously unrecognized link between the 3D flow topology of vortices and reaction hotspots. A novel chemiluminescence reaction was adopted to visualize the spatial map of reaction rates in channel intersections across a wide range of Reynolds numbers ($Re$). Further, flow and reactive transport simulations were experimentally cross-validated and used to demonstrate the role of 3D vortex topology on the emergence of reaction hotspots where reaction products are actively produced. To demonstrate the ubiquitous nature of vortex-induced reaction hotspots, we conducted experiments on rough-walled channels and also performed simulations over wide ranges of channel dimensions and Damk\"{o}hler numbers ($Da$). 

% methods - experiment
\paragraph{\label{sec:expt} Microfluidic experiment}
We conducted microfluidic experiments with chemiluminescence reaction~\citep{Jonsson1999} to visualize mixing and reaction at intersections. The mixing-induced reaction was performed by injecting two reactive solutions, labeled A and B, into two separate inlets on a polydimethylsiloxane (PDMS) microfluidic chip using a pulsation-free syringe pump (neMESYS 290N, Cetoni, Korbussen, Germany). The channels had a constant aperture of $100$ $\mu$m, a depth of $70$ $\mu$m, and a channel length of $2$ cm. The two channels intersected orthogonally at the center (1 cm) of their lengths at which the solutions mixed, and the chemiluminescence bimolecular reaction (A + B $\rightarrow$ C) occurred thereafter. 

A reaction between A and B produces a photon, and the produced photons were detected by a scientific CMOS camera (Orca-Flash4.0, Hamamatsu, Shizuoka, Japan) connected to a motorized inverted microscope system (TI2-E Nikon). The spatial map of reaction rate, $\frac{dc}{dt}$, was estimated by normalizing the accumulated light intensity values, which is proportional to $\Delta$c, by the exposure time, $\Delta$t~\citep{DeAnna2014a}. The composition of solution A was 1.5 mM of 1,8-diazabicyclo-[5,4,0]-undec-7-ene (DBU), 15 mM of 1,2,4-Triazole, 0.15 mM of 3- aminofluoranthen (3 – AFA), and 3 mM of H$_2$O$_2$. The composition of solution B was 3 mM of bis(2,4,6- trichlorophenyl)oxalate (TCPO). The solutes were dissolved in acetonitrile, and the experiments were performed at $25^\circ$C. All the chemicals were purchased from Sigma-Aldrich (MO, USA), and the details of the reaction mechanism are described in~\citet{Jonsson1999}. For passive tracer experiments, plain solvent and a solution containing 3 mM of 3 - AFA, which is a fluorescently active species, were separately injected into the two inlets, and the transport of the tracer was monitored via a green fluorescent protein filter (EX: 470/40nm, EM: 525/50nm). 

We investigated the inertia effects on the flow and reactive transport by varying $Re$ in the range of $1$ -- $300$, which commonly occur in natural and engineering processes~\citep{thompson1991effect,chun2006inertial,lee2015tail,nissan2018inertial,nissan2019Pe}. $Re$ was defined as $\frac{U_0 h}{\nu}$ where $U_0$ is the average flow velocity through a channel, $h$ is the aperture of the channel, and $\nu$ is the kinematic viscosity of the fluid. $Da$ is defined as $\frac{c_0 h^2 k}{D}$, where $D$ is the diffusion coefficient of solutes, $k$ is the reaction constant, and $c_0$ is the initial solute concentration. The experiments were conducted under seven different Reynolds numbers: $Re=[1,10,20,50,100,150,300]$. For all studied cases, both flow and concentration fields reach steady state. The estimated $Da$ in this study was $6.25$, and this implies that the system was relatively diffusion-limited with respect to the reaction.

%$Da$ in this study was $6.25$, and $Pe$ varies from $190$ to $56,000$. This implies that the system was relatively diffusion-limited with respect to the reaction, and the system was advection-dominant system.

%In addition, the $Pe$ and $Da$ were defined as $\frac{U_0}{Dh}$ and $\frac{c_0 h k}{U_0}$, respectively, where $D$ is the diffusion coefficient of solutes, $k$ is the reaction constant, and $c_0$ is the initial solute concentration. $Re$ of $1, 10, 20, 50, 100, 150,$ and $300$ were considered for the experiments. It should be noted that corresponding $Da$ for the low conditions considered in this study are calculated to be low ($ < 1$). This implies that upon the mixture of reactants, a portion of the reactants react instantaneously and unreacted ones are consumed in the downstream.

% methods - flow and reactive transport simulation
\paragraph{\label{sec:flowsim} Flow and reactive transport simulation}
We cross-validated experimental results with direct numerical simulations. The fluid flow simulations were performed in COMSOL Multiphysics (ver. 5.3). The density and kinematic viscosity of acetonitrile are $787$ kg/m$^3$ and $1.6\times10^{-6}$ m$^2$/s, respectively. The fluid flow was induced by setting a fixed inlet flow rate which determines $Re$, and the flow fields were obtained by solving the continuity equation and the Navier-Stokes equations with the finite element method. The flow channel domain were discretized into $1.8\times10^6$ elements for 3D simulations and into $5\times10^3$ elements for 2D simulations, and no slip boundary conditions were assigned at channel walls. 2D simulations assume parallel plate flows and neglect the boundary effects from the top and bottom boundaries. All of the flow simulations were converged to steady-state flow fields.

The flow field solutions were then coupled with the advection-diffusion-reaction equation~\citep{Dou2018}:

\begin{equation}
	\frac{\partial c_i}{\partial t}=\nabla  \cdot (uc_i) - \nabla \cdot(D_i \nabla c_i) + R_i
\end{equation}

where $c_i$ is the concentration of solute $i$, t is the time, $D_i$ is the diffusion coefficient of solute $i$, and $R_i$ is the reaction rate of solute $i$. The subscript $i$ represents species A, B, and C involved in the reaction. The limiting agents H$_2$O$_2$ and TCPO were chosen as the representative species for solutions A and B, respectively, and their initial concentrations of 3 mM were introduced into the two separate inlets. The diffusion coefficient of $3\times10^{-9}$ m$^2$/s was used for H$_2$O$_2$~\citep{Valencia2011}, and $1.6\times10^{-9}$ m$^2$/s was used for TCPO~\citep{DeAnna2014a} and product C. The temperature was set to $25^\circ$C in the model. The reaction between A and B is irreversible and the rate of loss of each reactant is equal to the rate of production of the product C which is described by a second-order reaction kinetics:

\begin{equation}
	R_i = \frac{dc_C}{dt}=-\frac{dc_A}{dt}=-\frac{dc_B}{dt}=kc_Ac_B
\end{equation}

where $k$ is the reaction constant defined as $k = \frac{1}{c_0\tau_r}$: $c_0$ is the initial solute concentration, and $\tau_r$ is the characteristic reaction time which is obtained experimentally~\citep{Connors1990,Jonsson1999,DeAnna2014a}. All of the transport simulations were converged to steady-state concentration fields.
%Previous studies have found the value to be $2$ s when the solute concentrations are one-third of the concentrations used in this study~\citep{Jonsson1999,DeAnna2014a}. Since the concentration difference was within one order, it was assumed that $\tau_r$ will not deviate significantly from $2$ s, thus, $\tau_r$ of $1$ s was assumed in this study. %The comparison between the microfluidics experiment and the simulation validates that 1 s was the proper choice.

% figure 1
\begin{figure}
\includegraphics[width=\columnwidth]{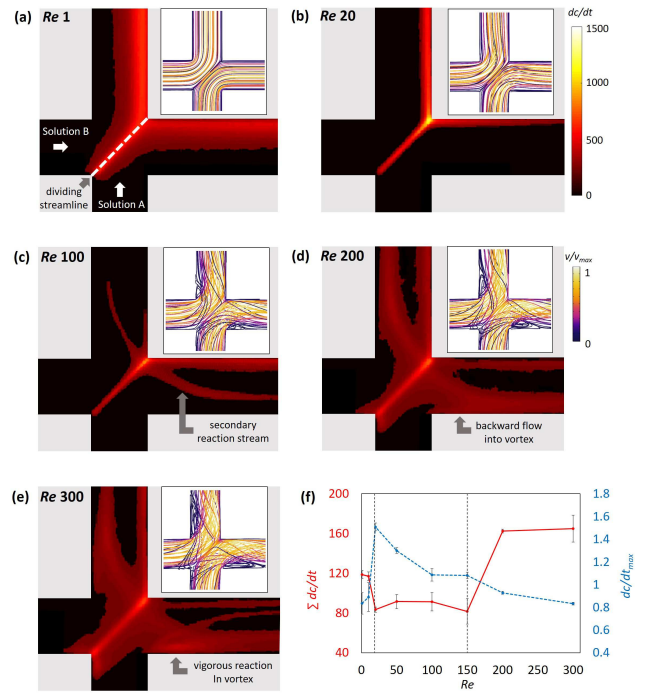}
\caption{\label{fig:expt} The spatial maps of reaction rate obtained from the microfluidic reaction experiments at $Re$ of (a) $1$, (b) $20$, (c) $100$, (d) $200$, and (e) $300$, and (f) the plots of total $\frac{dc}{dt}$ (red solid line) and maximum $\frac{dc}{dt}$ (blue dashed line). The color scale represents the light intensity divided by the exposure time which is proportional to the reaction rate, $\frac{dc}{dt}$. Insets: streamlines obtained from flow simulations with the color scale showing a normalized velocity magnitude.}
\end{figure}

% discussion on figure 1 - microfluidic exp't and flow sim.
\paragraph{\label{sec:expt&flow} Experimental observation of vortex-induced reaction hotspots}
The microfluidic experimental results from a straight orthogonal intersection are shown in Fig.~\ref{fig:expt}, and the streamlines obtained from flow simulations are shown in the insets. The spatial map of $\frac{dc}{dt}$ shows notable changes in the reaction dynamics as $Re$ increases from $1$ to $300$. Particularly, vortices seem to be strongly involved in the reaction at $Re$ greater than $200$. %The reaction dynamics near the intersection can be caterogrized into three regimes as discussed in the following. 

At $Re = 1$ (Fig.~\ref{fig:expt}(a)), the reaction occurs through a diffusive mixing of A and B along the dividing streamline and continues downstream. The analysis of streamlines confirmed that, across all $Re$, no streamlines enter the opposite stream, and the two inlet flows are separated along the dividing streamline. This implies that tracers can travel across the dividing streamline only by diffusion. Similar reaction dynamics are observed at $Re = 20$, but the width of the reaction band and the total reaction rate $\sum\frac{dc}{dt}$  decreased while the maximum intensity $\frac{dc}{dt}_{\text{max}}$ increased (Figs.~\ref{fig:expt}(b) and (f)). The increase in flow rate decreased the solute residence time, thereby reducing the amount of diffusive mixing. Consequently, the concentration gradient of solutes at the solution interface increased thereby elevating the local reaction rate (i.e., light intensity). On the other hand, the reduced reaction area and the solute residence time, collectively, lowered the total reaction rate $\sum\frac{dc}{dt}$ at $Re = 20$. 
%a broad band of light originating from the inner vertex of the intersection (indicated by the grey arrow in Fig.~\ref{fig:expt}a) extends diagonally along the solution interface to the opposite vertex and diverges out to two outlets.

At $Re=100$, a parabolic secondary reaction stream emerges from the interface (indicated by an arrow in Fig.~\ref{fig:expt}(c)). The streamlines in the inset show the emergence of twisting secondary flows around the corner. Such 3D helical streamlines in the direction of flow characterize a dean flow~\citep{Nivedita2017}, and the path of secondary reaction stream from the experiment was consistent with the dean flow streamlines obtained from the flow simulation. The secondary reaction stream increases the total reaction area and decreases the maximum reaction rate by disturbing the high concentration gradient along the dividing streamline. The decrease in the solute residence time and the maximum reaction rate from $Re=20$ to $150$ is balanced by the increase in the total reaction area leading to a relatively constant total reaction rate.
%This effectively dilutes the solutes at the solution interface thereby reducing the maximum reaction rate at $Re > 20$. However, the onset of the secondary reaction stream increases the reaction area. This 'tug-of-war' between decreasing maximum reaction rate and increasing reaction area is  balanced at $Re$ between $20$ and $150$ resulting in a relatively constant total reaction rate. Because the flow condition considered in this study is advection-dominant ($Pe > 1 \times 10^5$), the secondary reaction stream must have been emerged mainly due to the mixing induced by the dean flow. 

At $Re=200$, the width of the secondary reaction streams broadens significantly and they enter the vortices (Fig.~\ref{fig:expt}(d)). This is more evident at $Re = 300$ at which the circular flow pattern in the vortex zone is more pronounced and reflected in the $\frac{dc}{dt}$ map (Fig.~\ref{fig:expt}(e)). In this regime, the secondary reaction streams carrying reactive species are connected to vortices where the reactants are further mixed and reacted. Because flow velocities in the vortices are significantly smaller than those in the main flow as shown in the insets, the local $Da$ number is higher in the vortex-zone causing the vortex-zone to become a local reaction hotspot. The vortex-induced reaction significantly increases the total reaction rate near the intersection (Fig.~\ref{fig:expt}(f)). The vortices also exist at $Re = 100$ but not strong enough to bring the secondary reactive streams into vortices. This highlights the importance of the connected flow paths between the secondary reactive streams and vortices in the formation of vortex-induced reaction hotspots. One can conjecture that only a 3D flow effect can realize the connected flow paths, and this will be highlighted in the next section.%However, once such connected reaction streamlines are establishd, further increase in $Re$, for example from $200$ to $300$, decreases the average solute residence time and the $dc/dt_{max}$. This leads to a relatively constant total reaction rate. 

To summarize, there are three distinctive regimes for reaction dynamics as a function of $Re$ (shown by dashed vertical lines in Fig.~\ref{fig:expt}(f)). At $Re < 20$, the reaction is controlled by the diffusive mixing along the dividing streamline. At $20 < Re < 150$, the secondary reaction streams control the reaction dynamics. At $Re > 150$, the vortex-induced reaction hotspots control the reaction dynamics. Based on our observations, we hypothesize that the connected 3D flow paths from the secondary reaction streams to vortices induce reaction hotspots, which significantly raise the reaction rates in the third reaction regime. We validate our hypothesis by performing flow topology analysis and comparing experimental results between 2D and 3D simulations.

\begin{figure}
\includegraphics[width=\columnwidth]{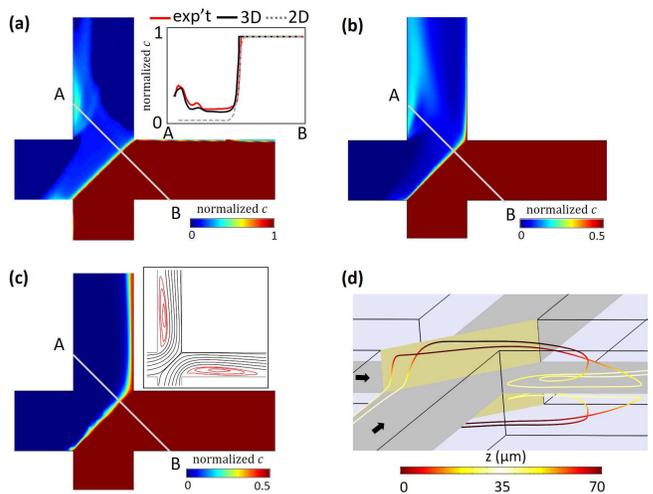}
\caption{\label{fig:trans} The projected spatial maps of tracer concentration at $Re$ = 300 obtained from (a) microfluidics experiment, (b) 3D simulation, and (c) 2D simulation. (d) The selected streamlines associated with vortex-connected streamlines. The yellow cross-surface shows the dividing stream surface (solution interface). The color bar indicates z-directional locations and highlights the z-directional motion of the spiral flow paths. Inset (a) Normalized projected concentration profiles along the cross-line AB. Inset (c) Streamlines obtained from the 2D simulation with red lines showing closed circular streamlines around the center type stagnation point.}
\end{figure}

% discussion on figure 2 - tracer test and transport sim.
\paragraph{\label{sec:trans} 3D vortex flow topology}
%To elucidate the flow topology of 3D vortices and their effects on tracer transport, 
We further studied transport characteristics by injecting a fluorescent passive tracer from the bottom inlet. Figure ~\ref{fig:trans}(a) shows the projected spatial map of tracer concentration obtained from the microfluidic experiment at $Re = 300$. The active transport of tracer from the dividing streamline to the vortex is clearly observed. The 2D projected tracer concentration map from the 3D simulation shows a very similar pattern with the experiment while the vortex in the 2D simulation has zero concentration (Figs.~\ref{fig:trans}(b) and (c)). Also, the experimental and 3D simulation results show multi-peak behavior which is not captured in 2D simulation (Fig.~\ref{fig:trans}(a) inset). 

The selected streamlines obtained from the 3D simulation (Fig.~\ref{fig:trans}(d)) reveal 3D spiral flow paths from the solution interface to the vortex. From the trajectories, we confirm that 3D vortices are formed at spiral saddle type stagnation points while the vortices in the 2D simulation are formed at center type stagnation points (Fig.~\ref{fig:trans}(c) inset). The 3D spiral flow paths advectively transport solutes from the solution interface to the vortex, but this is not possible in 2D vortices that do not have flow connectivity with the main flow paths. The general occurrence of spiral saddle stagnation points in 3D as opposed to center stagnation points in 2D is a fundamental difference between 2D and 3D flow topologies~\citep{Bresciani2019}. The 3D topology enables connectivity between main flow paths and vortices via 3D spiral flow paths, and this leads to the multi-peak behavior. 
% figure 2

\begin{figure}
\includegraphics[width=\columnwidth]{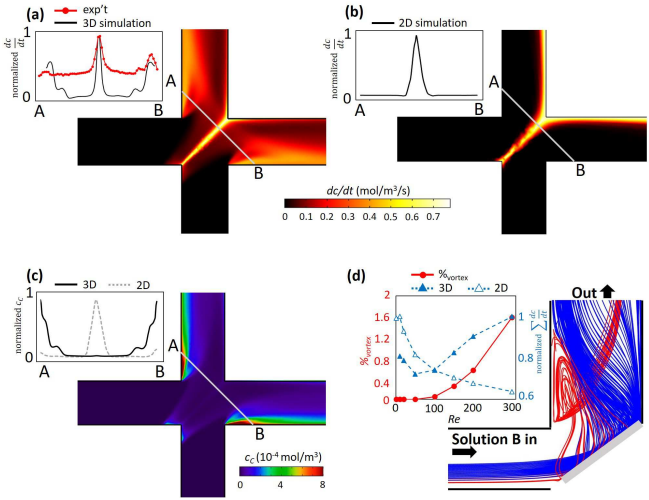}
\caption{\label{fig:reactivetrans} The projected spatial maps of local reaction rate, $dc_C/dt$ at $Re = 300$ obtained from (a) 3D simulation, and (b) 2D simulation. Insets: the $\frac{dc_C}{dt}$ profile along the cross-line AB shown with the grey line. (c) The reaction product concentration, $c_C$, from 3D simulation. Inset: the $c_C$ concentration profile along the cross-line AB for 3D and 2D simulations. (d) The illustration of reactive streamlines at $Re = 300$. The grey line shows a dividing streamline, and only a half of the intersection is shown because the system is symmetric. Inset: the plot of $\%_\text{vortex}$ and normalized total reaction rates as a function of $Re$.}
\end{figure}

% discussion on figure 3 - reactive transport sim.
\paragraph{\label{sec:reactivetrans} 3D vortex-induced reaction hotspots}
We performed reactive transport simulations to confirm 3D vortex-induced reaction hotspots. The projected spatial map of reaction rate, $\frac{dc_C}{dt}$, obtained from the 3D simulation is consistent with the experiment in which local reaction hotspots are formed at vortices (Fig.~\ref{fig:reactivetrans}(a)). On the other hand, the vortices in the 2D simulation are non-reactive (Fig.~\ref{fig:reactivetrans}(b)). This discrepancy is caused by the flow topology of 2D vortices that do not have flow connectivity with main flow paths (Fig.~\ref{fig:trans}(c) inset). Notably, not only is the reaction rate, $\frac{dc_C}{dt}$, high in the vortices, but the product concentration, $c_C$, also increases significantly towards the 3D vortices (Fig.~\ref{fig:reactivetrans}(c)). The lowered local velocity in the vortex zone allows the products to accumulate in the vortices. In contrast, the product concentration is maximum along the dividing streamline in the 2D simulation (Fig.~\ref{fig:reactivetrans}(c) inset). These results suggest that the 3D connected flow paths can turn vortices into reaction hotspots with not only high local reaction rates but also high product concentrations. This implies that for multi-species reactive systems, successive reactions involving reaction products will also actively occur in vortices.
%As a result, local peaks for $R$ emerges in the 3D vortex which indicate that 3D flow connected enables vortices to become local reactors 

% figure 3

%A close inspection of streamlines in Fig.~\ref{fig:reactivetrans}e clearly illustrates a vortex-induced local reaction hotspot. 
We now directly quantify the link between reaction and vortices. The streamlines that contain solutes from the opposite solution at a concentration greater than $0.01$ (i.e. $\frac{c_A}{c_0} > 0.01$) at least one point along their paths are defined as reactive streamlines. In other words, the reactive streamlines describe streamlines containing both reactants with concentrations greater than $0.01$. Among reactive streamlines, red streamlines indicate those that are drawn into a vortex while blue streamlines denote those that do not enter a vortex (Fig.~\ref{fig:reactivetrans}(d)). The pattern of red streamlines in Fig.~\ref{fig:reactivetrans}(d) is consistent with the reaction pattern obtained in the experiment (Fig.~\ref{fig:expt}(e)). This indicates that the flow connectivity between the reactive streams and vortices is critical in the generation of reaction hotspots.    
%Reactive streamlines are streamlines satisfying $\frac{c_A}{c_0} > 0.1$ at least one point on their path. Red streamlines indicate those that are drawn into a vortex while blue streamlines denote those that are not associated with the vortex. 
%To confirm that the rapid rise in total reaction rate observed in the experiments at $Re > 200$ is due to the connected flow path between the reactive stream and vortices, 

The connectedness of the reactive streamlines with vortices is quantified by calculating the percentage of the red streamlines with respect to the total reactive streamlines, i.e., $\%_\text{vortex}$. This percentage, $\%_\text{vortex}$, and the normalized total reaction rates obtained from the 3D and 2D simulations are plotted as a function of $Re$ (Fig.~\ref{fig:reactivetrans}(d) inset). The increase in $\%_\text{vortex}$ from $Re = 50$ strongly correlates with the increase in the total reaction rates in the 3D simulation. In contrast, the 2D simulation shows the opposite trend. This result indicates that a 3D description of flow and reaction at  intersections is essential to capture reaction dynamics. Although the degree of the connectedness of the vortex dramatically increases from $Re = 200$ to $Re = 300$, the total reaction rate does not exhibit a similar behavior. This result is consistent with the experiment (Fig.~\ref{fig:expt}(f)), and it is due to the increased local velocity in the vortices which decrease $Da$ inside of vortices. This confirms that both the 3D vortex flow topology and the decreased velocity in vortices are critical for initiating reaction hotspots.

% figure 4
\begin{figure}
\includegraphics[width=\columnwidth]{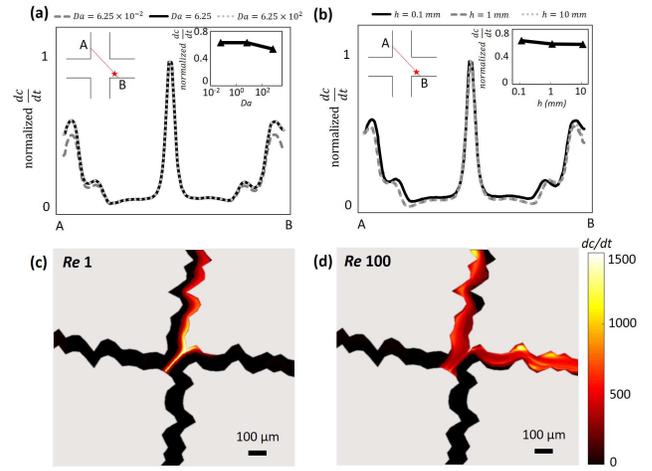}
\caption{\label{fig:rough} (a) The normalized reaction rate profiles along the cross-line AB as shown in the upper left inset. The red star in the inset shows the location of the maximum reaction rate. Upper right inset: the normalized maximum reaction rate in the vortex zone, $\frac{dc}{dt}_\text{max}$, obtained from 3D reactive transport simulations for a range of $Da$. (b) The normalized reaction rate profiles along the cross-line AB for a range of channel widths, $h$, from $100~\mu$m to $1$ cm. (c) Reaction rate maps obtained from microfluidic experiments with rough-walled microchannel intersection at $Re$ of $1$, and (d) $100$.} 
\end{figure}

% discussion on figure 4 - generality
\paragraph{\label{sec:generality} Generality.}
%The dimension of flow channels varies widely depending on the flow system. For example, industrial or household pipe network often consist of pipes having diameters in meters and centimeters while the scale of natural and physiological flow channels can range from micrometers to centimeters or even hundreds of meter for large river flows. To verify that the emergence of 3D vortex-induced reaction hotspots is scale-invariant, 3D reactive simulation with different values of channel aperture, $h$ = 1 m and 1 cm, is performed at $Re$ = 300. The result shows that the vortices formed at the intersection are reactive just as when $h$ is considered to be 100 $\mu m$ (Fig. ~\ref{fig:rough}a \& b). 

The size of flow channels can vary widely depending on the flow system, and the reaction rate can also vary widely depending on the reaction type leading to a wide range of $Da$ numbers. To study the generality of vortex-induced reaction hotspots, we conducted 3D simulations with different orders of channel aperture, $h = 1$ mm and $1$ cm, and $Da$ numbers of $0.01$ and $100$ at $Re = 300$. The depth of the channel was also changed to keep the same aperture to depth aspect ratio ($1:0.7$), and the $Da$ number was altered by changing the characteristic reaction time, $\tau_r$. The normalized reaction rate along the cross-line AB, and the maximum reaction rate in the vortex zone, $\frac{dc}{dt}_\text{max}$, are plotted in Figs.~\ref{fig:rough}(a) and (b). Regardless of channel dimension and reaction rates, we observe a ubiquitous nature of vortex-induced reaction hotspots.

The surfaces of flow channels are often rough, and the wall-roughness is known to promote the formation of vortices at lower $Re$, thereby impacting flow and transport~\citep{lee2015tail}. We performed experiments on a rough channel intersection to study the roughness effect on vortex-induced reaction hotspots. For generating rough surfaces, the Hurst exponent of $0.7$ was used~\citep{Dou2018,McDonald1974,Vigolo2014}. The channel had a constant aperture of $100~\mu$m and a depth of $70~\mu$m. The experiments were performed at $Re$ of $1$ and $100$. At $Re = 1$, the reaction occurred along the dividing streamline via diffusive mixing (Fig.~\ref{fig:rough}(c)). At $Re = 100$, the reaction pattern changed significantly due to the dean flow and 3D vortices formed at protruded areas (Fig.~\ref{fig:rough}(d)). Note that such 3D flow characteristics emerged at higher $Re$ in the straight intersection. This result implies that the vortex-induced reaction hotspots will more readily occur in rough-walled channel flows. 

% conclusion
In conclusion, we establish the mechanistic understanding of the vortex-induced reaction hotspots and their ubiquitous nature for the first time. 3D vortices occur at spiral saddle type stagnation points, and this 3D flow topology is essential in establishing the connected flow paths from the mainstream to vortices through which the reactants can enter the vortices advectively. In addition, the increased solute residence time inside the vortices due to the lower flow velocity, compared to the main flow, facilitated the formation of a vortex-induced reaction hotspot. Vortex-induced reaction hotspots are shown to occur over a wide range of channel dimensions and reaction rates, and they become more vigorous in rough channels. These results have direct implications in many engineering and natural processes involving mixing and reaction in channel flows.  \\
%which has direct implications to many reactive systems.
%This phenomenon is not considered in the conventional solute mixing models (e.g. streamline routing and complete mixing) in which monotonic decay of solute is anticipated, and thus, cannot consider the effects of the multi-peak behavior. 3D effect leads to very distinct reaction dynamics and spatial reaction patterns.
%As such, the influence of vortex-induced reaction hotspots should not be neglected when dealing with various reactive transport in rough flow channels.  
%Hence, 3D description of reactive transport is crucial for accurately representing reactions in microchannel systems. 
%Outlook and impact: precipitation or dissolution because the vortex zone will be highly subceptible to change in channel geometry.
%\subsection{Acknowledgments}

We sincerely appreciate Dr. Etienne Bresicani for the helpful discussions on flow topology. The authors thank the Minnesota Supercomputing Institute (MSI) at the University of Minnesota for computational resources, and also acknowledge support from NSF via a grant EAR1813526. PKK acknowledges the College of Science \& Engineering at the University of Minnesota and the George and Orpha Gibson Endowment. 

\bibliographystyle{apsrev4-1}

\bibliography{manuscript_lee_kang}% Produces the bibliography via BibTeX.

\begin{thebibliography}{33}%
\makeatletter
\providecommand \@ifxundefined [1]{%
 \@ifx{#1\undefined}
}%
\providecommand \@ifnum [1]{%
 \ifnum #1\expandafter \@firstoftwo
 \else \expandafter \@secondoftwo
 \fi
}%
\providecommand \@ifx [1]{%
 \ifx #1\expandafter \@firstoftwo
 \else \expandafter \@secondoftwo
 \fi
}%
\providecommand \natexlab [1]{#1}%
\providecommand \enquote  [1]{``#1''}%
\providecommand \bibnamefont  [1]{#1}%
\providecommand \bibfnamefont [1]{#1}%
\providecommand \citenamefont [1]{#1}%
\providecommand \href@noop [0]{\@secondoftwo}%
\providecommand \href [0]{\begingroup \@sanitize@url \@href}%
\providecommand \@href[1]{\@@startlink{#1}\@@href}%
\providecommand \@@href[1]{\endgroup#1\@@endlink}%
\providecommand \@sanitize@url [0]{\catcode `\\12\catcode `\$12\catcode
  `\&12\catcode `\#12\catcode `\^12\catcode `\_12\catcode `\%12\relax}%
\providecommand \@@startlink[1]{}%
\providecommand \@@endlink[0]{}%
\providecommand \url  [0]{\begingroup\@sanitize@url \@url }%
\providecommand \@url [1]{\endgroup\@href {#1}{\urlprefix }}%
\providecommand \urlprefix  [0]{URL }%
\providecommand \Eprint [0]{\href }%
\providecommand \doibase [0]{http://dx.doi.org/}%
\providecommand \selectlanguage [0]{\@gobble}%
\providecommand \bibinfo  [0]{\@secondoftwo}%
\providecommand \bibfield  [0]{\@secondoftwo}%
\providecommand \translation [1]{[#1]}%
\providecommand \BibitemOpen [0]{}%
\providecommand \bibitemStop [0]{}%
\providecommand \bibitemNoStop [0]{.\EOS\space}%
\providecommand \EOS [0]{\spacefactor3000\relax}%
\providecommand \BibitemShut  [1]{\csname bibitem#1\endcsname}%
\let\auto@bib@innerbib\@empty
%</preamble>
\bibitem [{\citenamefont {Kosakowski}\ and\ \citenamefont
  {Berkowitz}(1999)}]{kosakowski1999}%
  \BibitemOpen
  \bibfield  {author} {\bibinfo {author} {\bibfnamefont {G.}~\bibnamefont
  {Kosakowski}}\ and\ \bibinfo {author} {\bibfnamefont {B.}~\bibnamefont
  {Berkowitz}},\ }\href@noop {} {\bibfield  {journal} {\bibinfo  {journal}
  {Geophys. Res. Lett.}\ }\textbf {\bibinfo {volume} {26}},\ \bibinfo {pages}
  {1765} (\bibinfo {year} {1999})}\BibitemShut {NoStop}%
\bibitem [{\citenamefont {Boutt}\ \emph {et~al.}(2006)\citenamefont {Boutt},
  \citenamefont {Grasselli}, \citenamefont {Fredrich}, \citenamefont {Cook},\
  and\ \citenamefont {Williams}}]{boutt2006trapping}%
  \BibitemOpen
  \bibfield  {author} {\bibinfo {author} {\bibfnamefont {D.~F.}\ \bibnamefont
  {Boutt}}, \bibinfo {author} {\bibfnamefont {G.}~\bibnamefont {Grasselli}},
  \bibinfo {author} {\bibfnamefont {J.~T.}\ \bibnamefont {Fredrich}}, \bibinfo
  {author} {\bibfnamefont {B.~K.}\ \bibnamefont {Cook}}, \ and\ \bibinfo
  {author} {\bibfnamefont {J.~R.}\ \bibnamefont {Williams}},\ }\href@noop {}
  {\bibfield  {journal} {\bibinfo  {journal} {Geophys. Res. Lett.}\ }\textbf
  {\bibinfo {volume} {33}} (\bibinfo {year} {2006})}\BibitemShut {NoStop}%
\bibitem [{\citenamefont {Cardenas}\ \emph {et~al.}(2009)\citenamefont
  {Cardenas}, \citenamefont {Slottke}, \citenamefont {Ketcham},\ and\
  \citenamefont {Sharp}}]{cardenas2009effects}%
  \BibitemOpen
  \bibfield  {author} {\bibinfo {author} {\bibfnamefont {M.~B.}\ \bibnamefont
  {Cardenas}}, \bibinfo {author} {\bibfnamefont {D.~T.}\ \bibnamefont
  {Slottke}}, \bibinfo {author} {\bibfnamefont {R.~A.}\ \bibnamefont
  {Ketcham}}, \ and\ \bibinfo {author} {\bibfnamefont {J.~M.}\ \bibnamefont
  {Sharp}},\ }\href@noop {} {\bibfield  {journal} {\bibinfo  {journal} {J.
  Geophys. Res. Solid Earth}\ }\textbf {\bibinfo {volume} {114}} (\bibinfo
  {year} {2009})}\BibitemShut {NoStop}%
\bibitem [{\citenamefont {Lee}\ \emph {et~al.}(2015)\citenamefont {Lee},
  \citenamefont {Yeo}, \citenamefont {Lee},\ and\ \citenamefont
  {Detwiler}}]{lee2015tail}%
  \BibitemOpen
  \bibfield  {author} {\bibinfo {author} {\bibfnamefont {S.~H.}\ \bibnamefont
  {Lee}}, \bibinfo {author} {\bibfnamefont {I.~W.}\ \bibnamefont {Yeo}},
  \bibinfo {author} {\bibfnamefont {K.-K.}\ \bibnamefont {Lee}}, \ and\
  \bibinfo {author} {\bibfnamefont {R.~L.}\ \bibnamefont {Detwiler}},\
  }\href@noop {} {\bibfield  {journal} {\bibinfo  {journal} {Geophys. Res.
  Lett.}\ }\textbf {\bibinfo {volume} {42}},\ \bibinfo {pages} {6340} (\bibinfo
  {year} {2015})}\BibitemShut {NoStop}%
\bibitem [{\citenamefont {Wood}(2007)}]{Wood2007}%
  \BibitemOpen
  \bibfield  {author} {\bibinfo {author} {\bibfnamefont {B.~D.}\ \bibnamefont
  {Wood}},\ }\href {\doibase 10.1029/2006WR005790} {\bibfield  {journal}
  {\bibinfo  {journal} {Water Resour. Res.}\ }\textbf {\bibinfo {volume}
  {43}},\ \bibinfo {pages} {1} (\bibinfo {year} {2007})}\BibitemShut {NoStop}%
\bibitem [{\citenamefont {Ye}\ \emph {et~al.}(2015)\citenamefont {Ye},
  \citenamefont {Chiogna}, \citenamefont {Cirpka}, \citenamefont {Grathwohl},\
  and\ \citenamefont {Rolle}}]{ye2015experimental}%
  \BibitemOpen
  \bibfield  {author} {\bibinfo {author} {\bibfnamefont {Y.}~\bibnamefont
  {Ye}}, \bibinfo {author} {\bibfnamefont {G.}~\bibnamefont {Chiogna}},
  \bibinfo {author} {\bibfnamefont {O.~A.}\ \bibnamefont {Cirpka}}, \bibinfo
  {author} {\bibfnamefont {P.}~\bibnamefont {Grathwohl}}, \ and\ \bibinfo
  {author} {\bibfnamefont {M.}~\bibnamefont {Rolle}},\ }\href@noop {}
  {\bibfield  {journal} {\bibinfo  {journal} {Phys. Rev. Lett.}\ }\textbf
  {\bibinfo {volume} {115}},\ \bibinfo {pages} {194502} (\bibinfo {year}
  {2015})}\BibitemShut {NoStop}%
\bibitem [{\citenamefont {Crevacore}\ \emph {et~al.}(2016)\citenamefont
  {Crevacore}, \citenamefont {Tosco}, \citenamefont {Sethi}, \citenamefont
  {Boccardo},\ and\ \citenamefont {Marchisio}}]{crevacore2016recirculation}%
  \BibitemOpen
  \bibfield  {author} {\bibinfo {author} {\bibfnamefont {E.}~\bibnamefont
  {Crevacore}}, \bibinfo {author} {\bibfnamefont {T.}~\bibnamefont {Tosco}},
  \bibinfo {author} {\bibfnamefont {R.}~\bibnamefont {Sethi}}, \bibinfo
  {author} {\bibfnamefont {G.}~\bibnamefont {Boccardo}}, \ and\ \bibinfo
  {author} {\bibfnamefont {D.~L.}\ \bibnamefont {Marchisio}},\ }\href@noop {}
  {\bibfield  {journal} {\bibinfo  {journal} {Phys. Rev. E}\ }\textbf {\bibinfo
  {volume} {94}},\ \bibinfo {pages} {053118} (\bibinfo {year}
  {2016})}\BibitemShut {NoStop}%
\bibitem [{\citenamefont {Pasquier}\ \emph {et~al.}(2017)\citenamefont
  {Pasquier}, \citenamefont {Quintard},\ and\ \citenamefont
  {Davit}}]{pasquier2017}%
  \BibitemOpen
  \bibfield  {author} {\bibinfo {author} {\bibfnamefont {S.}~\bibnamefont
  {Pasquier}}, \bibinfo {author} {\bibfnamefont {M.}~\bibnamefont {Quintard}},
  \ and\ \bibinfo {author} {\bibfnamefont {Y.}~\bibnamefont {Davit}},\
  }\href@noop {} {\bibfield  {journal} {\bibinfo  {journal} {Chem. Eng.}\
  }\textbf {\bibinfo {volume} {165}},\ \bibinfo {pages} {131} (\bibinfo {year}
  {2017})}\BibitemShut {NoStop}%
\bibitem [{\citenamefont {Ault}\ \emph {et~al.}(2016)\citenamefont {Ault},
  \citenamefont {Fani}, \citenamefont {Chen}, \citenamefont {Shin},
  \citenamefont {Gallaire},\ and\ \citenamefont {Stone}}]{Ault2016}%
  \BibitemOpen
  \bibfield  {author} {\bibinfo {author} {\bibfnamefont {J.~T.}\ \bibnamefont
  {Ault}}, \bibinfo {author} {\bibfnamefont {A.}~\bibnamefont {Fani}}, \bibinfo
  {author} {\bibfnamefont {K.~K.}\ \bibnamefont {Chen}}, \bibinfo {author}
  {\bibfnamefont {S.}~\bibnamefont {Shin}}, \bibinfo {author} {\bibfnamefont
  {F.}~\bibnamefont {Gallaire}}, \ and\ \bibinfo {author} {\bibfnamefont
  {H.~A.}\ \bibnamefont {Stone}},\ }\href {\doibase
  10.1103/PhysRevLett.117.084501} {\bibfield  {journal} {\bibinfo  {journal}
  {Phys. Rev. Lett.}\ }\textbf {\bibinfo {volume} {117}},\ \bibinfo {pages}
  {084501} (\bibinfo {year} {2016})}\BibitemShut {NoStop}%
\bibitem [{\citenamefont {Oettinger}\ \emph {et~al.}(2018)\citenamefont
  {Oettinger}, \citenamefont {Ault}, \citenamefont {Stone},\ and\ \citenamefont
  {Haller}}]{oettinger2018}%
  \BibitemOpen
  \bibfield  {author} {\bibinfo {author} {\bibfnamefont {D.}~\bibnamefont
  {Oettinger}}, \bibinfo {author} {\bibfnamefont {J.~T.}\ \bibnamefont {Ault}},
  \bibinfo {author} {\bibfnamefont {H.~A.}\ \bibnamefont {Stone}}, \ and\
  \bibinfo {author} {\bibfnamefont {G.}~\bibnamefont {Haller}},\ }\href@noop {}
  {\bibfield  {journal} {\bibinfo  {journal} {Phys. Rev. Lett.}\ }\textbf
  {\bibinfo {volume} {121}},\ \bibinfo {pages} {054502} (\bibinfo {year}
  {2018})}\BibitemShut {NoStop}%
\bibitem [{\citenamefont {Xi}\ \emph {et~al.}(2004)\citenamefont {Xi},
  \citenamefont {Marks}, \citenamefont {Parikh}, \citenamefont {Raskin},\ and\
  \citenamefont {Boppart}}]{Xi7516}%
  \BibitemOpen
  \bibfield  {author} {\bibinfo {author} {\bibfnamefont {C.}~\bibnamefont
  {Xi}}, \bibinfo {author} {\bibfnamefont {D.~L.}\ \bibnamefont {Marks}},
  \bibinfo {author} {\bibfnamefont {D.~S.}\ \bibnamefont {Parikh}}, \bibinfo
  {author} {\bibfnamefont {L.}~\bibnamefont {Raskin}}, \ and\ \bibinfo {author}
  {\bibfnamefont {S.~A.}\ \bibnamefont {Boppart}},\ }\href {\doibase
  10.1073/pnas.0402433101} {\bibfield  {journal} {\bibinfo  {journal} {Proc.
  Natl. Acad. Sci.}\ }\textbf {\bibinfo {volume} {101}},\ \bibinfo {pages}
  {7516} (\bibinfo {year} {2004})}\BibitemShut {NoStop}%
\bibitem [{\citenamefont {Gharib}\ \emph {et~al.}(2006)\citenamefont {Gharib},
  \citenamefont {Rambod}, \citenamefont {Kheradvar}, \citenamefont {Sahn},\
  and\ \citenamefont {Dabiri}}]{Gharib2006}%
  \BibitemOpen
  \bibfield  {author} {\bibinfo {author} {\bibfnamefont {M.}~\bibnamefont
  {Gharib}}, \bibinfo {author} {\bibfnamefont {E.}~\bibnamefont {Rambod}},
  \bibinfo {author} {\bibfnamefont {A.}~\bibnamefont {Kheradvar}}, \bibinfo
  {author} {\bibfnamefont {D.~J.}\ \bibnamefont {Sahn}}, \ and\ \bibinfo
  {author} {\bibfnamefont {J.~O.}\ \bibnamefont {Dabiri}},\ }\href@noop {}
  {\bibfield  {journal} {\bibinfo  {journal} {Proc. Natl. Acad. Sci.}\ }\textbf
  {\bibinfo {volume} {103}},\ \bibinfo {pages} {6305} (\bibinfo {year}
  {2006})}\BibitemShut {NoStop}%
\bibitem [{\citenamefont {Biasetti}\ \emph {et~al.}(2011)\citenamefont
  {Biasetti}, \citenamefont {Hussain},\ and\ \citenamefont {{Christian
  Gasser}}}]{Biasetti2011}%
  \BibitemOpen
  \bibfield  {author} {\bibinfo {author} {\bibfnamefont {J.}~\bibnamefont
  {Biasetti}}, \bibinfo {author} {\bibfnamefont {F.}~\bibnamefont {Hussain}}, \
  and\ \bibinfo {author} {\bibfnamefont {T.}~\bibnamefont {{Christian
  Gasser}}},\ }\href {\doibase 10.1098/rsif.2011.0041} {\bibfield  {journal}
  {\bibinfo  {journal} {J. R. Soc. Interface}\ }\textbf {\bibinfo {volume}
  {8}},\ \bibinfo {pages} {1449} (\bibinfo {year} {2011})}\BibitemShut
  {NoStop}%
\bibitem [{\citenamefont {D{\'e}lery}(2001)}]{delery2001}%
  \BibitemOpen
  \bibfield  {author} {\bibinfo {author} {\bibfnamefont {J.~M.}\ \bibnamefont
  {D{\'e}lery}},\ }\href@noop {} {\bibfield  {journal} {\bibinfo  {journal}
  {Annu. Rev. Fluid Mech.}\ }\textbf {\bibinfo {volume} {33}},\ \bibinfo
  {pages} {129} (\bibinfo {year} {2001})}\BibitemShut {NoStop}%
\bibitem [{\citenamefont {Haller}(2001)}]{haller2001}%
  \BibitemOpen
  \bibfield  {author} {\bibinfo {author} {\bibfnamefont {G.}~\bibnamefont
  {Haller}},\ }\href@noop {} {\bibfield  {journal} {\bibinfo  {journal}
  {Physica D}\ }\textbf {\bibinfo {volume} {149}},\ \bibinfo {pages} {248}
  (\bibinfo {year} {2001})}\BibitemShut {NoStop}%
\bibitem [{\citenamefont {Bresciani}\ \emph {et~al.}(2019)\citenamefont
  {Bresciani}, \citenamefont {Kang},\ and\ \citenamefont
  {Lee}}]{Bresciani2019}%
  \BibitemOpen
  \bibfield  {author} {\bibinfo {author} {\bibfnamefont {E.}~\bibnamefont
  {Bresciani}}, \bibinfo {author} {\bibfnamefont {P.~K.}\ \bibnamefont {Kang}},
  \ and\ \bibinfo {author} {\bibfnamefont {S.}~\bibnamefont {Lee}},\ }\href
  {\doibase 10.1029/2018WR023508} {\bibfield  {journal} {\bibinfo  {journal}
  {Water Resour. Res.}\ }\textbf {\bibinfo {volume} {55}},\ \bibinfo {pages}
  {1624} (\bibinfo {year} {2019})}\BibitemShut {NoStop}%
\bibitem [{\citenamefont {de~Barros}\ \emph {et~al.}(2012)\citenamefont
  {de~Barros}, \citenamefont {Dentz}, \citenamefont {Koch},\ and\ \citenamefont
  {Nowak}}]{de2012flow}%
  \BibitemOpen
  \bibfield  {author} {\bibinfo {author} {\bibfnamefont {F.~P.}\ \bibnamefont
  {de~Barros}}, \bibinfo {author} {\bibfnamefont {M.}~\bibnamefont {Dentz}},
  \bibinfo {author} {\bibfnamefont {J.}~\bibnamefont {Koch}}, \ and\ \bibinfo
  {author} {\bibfnamefont {W.}~\bibnamefont {Nowak}},\ }\href@noop {}
  {\bibfield  {journal} {\bibinfo  {journal} {Geophys. Res. Lett.}\ }\textbf
  {\bibinfo {volume} {39}} (\bibinfo {year} {2012})}\BibitemShut {NoStop}%
\bibitem [{\citenamefont {Engdahl}\ \emph {et~al.}(2014)\citenamefont
  {Engdahl}, \citenamefont {Benson},\ and\ \citenamefont
  {Bolster}}]{engdahl2014predicting}%
  \BibitemOpen
  \bibfield  {author} {\bibinfo {author} {\bibfnamefont {N.~B.}\ \bibnamefont
  {Engdahl}}, \bibinfo {author} {\bibfnamefont {D.~A.}\ \bibnamefont {Benson}},
  \ and\ \bibinfo {author} {\bibfnamefont {D.}~\bibnamefont {Bolster}},\
  }\href@noop {} {\bibfield  {journal} {\bibinfo  {journal} {Phys. Rev. E}\
  }\textbf {\bibinfo {volume} {90}},\ \bibinfo {pages} {051001} (\bibinfo
  {year} {2014})}\BibitemShut {NoStop}%
\bibitem [{\citenamefont {Turuban}\ \emph {et~al.}(2018)\citenamefont
  {Turuban}, \citenamefont {Lester}, \citenamefont {Le~Borgne},\ and\
  \citenamefont {M{\'e}heust}}]{turuban2018space}%
  \BibitemOpen
  \bibfield  {author} {\bibinfo {author} {\bibfnamefont {R.}~\bibnamefont
  {Turuban}}, \bibinfo {author} {\bibfnamefont {D.~R.}\ \bibnamefont {Lester}},
  \bibinfo {author} {\bibfnamefont {T.}~\bibnamefont {Le~Borgne}}, \ and\
  \bibinfo {author} {\bibfnamefont {Y.}~\bibnamefont {M{\'e}heust}},\
  }\href@noop {} {\bibfield  {journal} {\bibinfo  {journal} {Phys. Rev. Lett.}\
  }\textbf {\bibinfo {volume} {120}},\ \bibinfo {pages} {024501} (\bibinfo
  {year} {2018})}\BibitemShut {NoStop}%
\bibitem [{\citenamefont {Lee}\ \emph {et~al.}(2016)\citenamefont {Lee},
  \citenamefont {Wang}, \citenamefont {Liu},\ and\ \citenamefont
  {Fu}}]{lee2016passive}%
  \BibitemOpen
  \bibfield  {author} {\bibinfo {author} {\bibfnamefont {C.-Y.}\ \bibnamefont
  {Lee}}, \bibinfo {author} {\bibfnamefont {W.-T.}\ \bibnamefont {Wang}},
  \bibinfo {author} {\bibfnamefont {C.-C.}\ \bibnamefont {Liu}}, \ and\
  \bibinfo {author} {\bibfnamefont {L.-M.}\ \bibnamefont {Fu}},\ }\href@noop {}
  {\bibfield  {journal} {\bibinfo  {journal} {Chem. Eng. J.}\ }\textbf
  {\bibinfo {volume} {288}},\ \bibinfo {pages} {146} (\bibinfo {year}
  {2016})}\BibitemShut {NoStop}%
\bibitem [{\citenamefont {Zou}\ \emph {et~al.}(2017)\citenamefont {Zou},
  \citenamefont {Jing},\ and\ \citenamefont {Cvetkovic}}]{Zou2017}%
  \BibitemOpen
  \bibfield  {author} {\bibinfo {author} {\bibfnamefont {L.}~\bibnamefont
  {Zou}}, \bibinfo {author} {\bibfnamefont {L.}~\bibnamefont {Jing}}, \ and\
  \bibinfo {author} {\bibfnamefont {V.}~\bibnamefont {Cvetkovic}},\ }\href
  {\doibase 10.1016/J.ADVWATRES.2017.06.003} {\bibfield  {journal} {\bibinfo
  {journal} {Adv. Water Resour.}\ }\textbf {\bibinfo {volume} {107}},\ \bibinfo
  {pages} {1} (\bibinfo {year} {2017})}\BibitemShut {NoStop}%
\bibitem [{\citenamefont {Jonsson}\ and\ \citenamefont
  {Irgum}(1999)}]{Jonsson1999}%
  \BibitemOpen
  \bibfield  {author} {\bibinfo {author} {\bibfnamefont {T.}~\bibnamefont
  {Jonsson}}\ and\ \bibinfo {author} {\bibfnamefont {K.}~\bibnamefont
  {Irgum}},\ }\href {\doibase 10.1016/S0003-2670(99)00626-1} {\bibfield
  {journal} {\bibinfo  {journal} {Anal. Chim. Acta}\ }\textbf {\bibinfo
  {volume} {400}},\ \bibinfo {pages} {257} (\bibinfo {year}
  {1999})}\BibitemShut {NoStop}%
\bibitem [{\citenamefont {{de Anna}}\ \emph {et~al.}(2014)\citenamefont {{de
  Anna}}, \citenamefont {Jimenez-Martinez}, \citenamefont {Tabuteau},
  \citenamefont {Turuban}, \citenamefont {{Le Borgne}}, \citenamefont
  {Derrien},\ and\ \citenamefont {M{\'{e}}heust}}]{DeAnna2014a}%
  \BibitemOpen
  \bibfield  {author} {\bibinfo {author} {\bibfnamefont {P.}~\bibnamefont {{de
  Anna}}}, \bibinfo {author} {\bibfnamefont {J.}~\bibnamefont
  {Jimenez-Martinez}}, \bibinfo {author} {\bibfnamefont {H.}~\bibnamefont
  {Tabuteau}}, \bibinfo {author} {\bibfnamefont {R.}~\bibnamefont {Turuban}},
  \bibinfo {author} {\bibfnamefont {T.}~\bibnamefont {{Le Borgne}}}, \bibinfo
  {author} {\bibfnamefont {M.}~\bibnamefont {Derrien}}, \ and\ \bibinfo
  {author} {\bibfnamefont {Y.}~\bibnamefont {M{\'{e}}heust}},\ }\href {\doibase
  10.1021/es403105b} {\bibfield  {journal} {\bibinfo  {journal} {Environ. Sci.
  Technol.}\ }\textbf {\bibinfo {volume} {48}},\ \bibinfo {pages} {508}
  (\bibinfo {year} {2014})}\BibitemShut {NoStop}%
\bibitem [{\citenamefont {Thompson}\ and\ \citenamefont
  {Brown}(1991)}]{thompson1991effect}%
  \BibitemOpen
  \bibfield  {author} {\bibinfo {author} {\bibfnamefont {M.~E.}\ \bibnamefont
  {Thompson}}\ and\ \bibinfo {author} {\bibfnamefont {S.~R.}\ \bibnamefont
  {Brown}},\ }\href@noop {} {\bibfield  {journal} {\bibinfo  {journal} {J.
  Geophys. Res. Solid Earth}\ }\textbf {\bibinfo {volume} {96}},\ \bibinfo
  {pages} {21923} (\bibinfo {year} {1991})}\BibitemShut {NoStop}%
\bibitem [{\citenamefont {Chun}\ and\ \citenamefont
  {Ladd}(2006)}]{chun2006inertial}%
  \BibitemOpen
  \bibfield  {author} {\bibinfo {author} {\bibfnamefont {B.}~\bibnamefont
  {Chun}}\ and\ \bibinfo {author} {\bibfnamefont {A.}~\bibnamefont {Ladd}},\
  }\href@noop {} {\bibfield  {journal} {\bibinfo  {journal} {Phys. Fluids}\
  }\textbf {\bibinfo {volume} {18}},\ \bibinfo {pages} {031704} (\bibinfo
  {year} {2006})}\BibitemShut {NoStop}%
\bibitem [{\citenamefont {Nissan}\ and\ \citenamefont
  {Berkowitz}(2018)}]{nissan2018inertial}%
  \BibitemOpen
  \bibfield  {author} {\bibinfo {author} {\bibfnamefont {A.}~\bibnamefont
  {Nissan}}\ and\ \bibinfo {author} {\bibfnamefont {B.}~\bibnamefont
  {Berkowitz}},\ }\href {\doibase 10.1103/PhysRevLett.120.054504} {\bibfield
  {journal} {\bibinfo  {journal} {Phys. Rev. Lett.}\ }\textbf {\bibinfo
  {volume} {120}},\ \bibinfo {pages} {054504} (\bibinfo {year}
  {2018})}\BibitemShut {NoStop}%
\bibitem [{\citenamefont {Nissan}\ and\ \citenamefont
  {Berkowitz}(2019)}]{nissan2019Pe}%
  \BibitemOpen
  \bibfield  {author} {\bibinfo {author} {\bibfnamefont {A.}~\bibnamefont
  {Nissan}}\ and\ \bibinfo {author} {\bibfnamefont {B.}~\bibnamefont
  {Berkowitz}},\ }\href {\doibase 10.1103/PhysRevE.99.033108} {\bibfield
  {journal} {\bibinfo  {journal} {Phys. Rev. E}\ }\textbf {\bibinfo {volume}
  {99}},\ \bibinfo {pages} {033108} (\bibinfo {year} {2019})}\BibitemShut
  {NoStop}%
\bibitem [{\citenamefont {Dou}\ \emph {et~al.}(2018)\citenamefont {Dou},
  \citenamefont {Zhou}, \citenamefont {Wang},\ and\ \citenamefont
  {Liu}}]{Dou2018}%
  \BibitemOpen
  \bibfield  {author} {\bibinfo {author} {\bibfnamefont {Z.}~\bibnamefont
  {Dou}}, \bibinfo {author} {\bibfnamefont {Z.}~\bibnamefont {Zhou}}, \bibinfo
  {author} {\bibfnamefont {J.}~\bibnamefont {Wang}}, \ and\ \bibinfo {author}
  {\bibfnamefont {J.}~\bibnamefont {Liu}},\ }\href {\doibase
  10.1155/2018/9095143} {\bibfield  {journal} {\bibinfo  {journal} {Geofluids}\
  }\textbf {\bibinfo {volume} {2018}},\ \bibinfo {pages} {1} (\bibinfo {year}
  {2018})}\BibitemShut {NoStop}%
\bibitem [{\citenamefont {Valencia}\ and\ \citenamefont
  {Gonz{\'{a}}lez}(2011)}]{Valencia2011}%
  \BibitemOpen
  \bibfield  {author} {\bibinfo {author} {\bibfnamefont {D.~P.}\ \bibnamefont
  {Valencia}}\ and\ \bibinfo {author} {\bibfnamefont {F.~J.}\ \bibnamefont
  {Gonz{\'{a}}lez}},\ }\href {\doibase 10.1016/j.elecom.2010.11.032} {\bibfield
   {journal} {\bibinfo  {journal} {Electrochem. commun.}\ }\textbf {\bibinfo
  {volume} {13}},\ \bibinfo {pages} {129} (\bibinfo {year} {2011})}\BibitemShut
  {NoStop}%
\bibitem [{\citenamefont {Connors}(1990)}]{Connors1990}%
  \BibitemOpen
  \bibfield  {author} {\bibinfo {author} {\bibfnamefont {K.~A.}\ \bibnamefont
  {Connors}},\ }\href@noop {} {\emph {\bibinfo {title} {Chemical kinetics: the
  study of reaction rates in solution}}}\ (\bibinfo  {publisher} {VCH},\
  \bibinfo {year} {1990})\ p.\ \bibinfo {pages} {480}\BibitemShut {NoStop}%
\bibitem [{\citenamefont {Nivedita}\ \emph {et~al.}(2017)\citenamefont
  {Nivedita}, \citenamefont {Ligrani},\ and\ \citenamefont
  {Papautsky}}]{Nivedita2017}%
  \BibitemOpen
  \bibfield  {author} {\bibinfo {author} {\bibfnamefont {N.}~\bibnamefont
  {Nivedita}}, \bibinfo {author} {\bibfnamefont {P.}~\bibnamefont {Ligrani}}, \
  and\ \bibinfo {author} {\bibfnamefont {I.}~\bibnamefont {Papautsky}},\ }\href
  {\doibase 10.1038/srep44072} {\bibfield  {journal} {\bibinfo  {journal} {Sci.
  Rep.}\ }\textbf {\bibinfo {volume} {7}},\ \bibinfo {pages} {44072} (\bibinfo
  {year} {2017})}\BibitemShut {NoStop}%
\bibitem [{\citenamefont {McDonald}(1974)}]{McDonald1974}%
  \BibitemOpen
  \bibfield  {author} {\bibinfo {author} {\bibfnamefont {D.~A.}\ \bibnamefont
  {McDonald}},\ }\href {\doibase 10.1146/annurev.fluid.29.1.399} {\bibfield
  {journal} {\bibinfo  {journal} {Annu. Rev. Fluid Mech.}\ }\textbf {\bibinfo
  {volume} {{\pounds}12.-}},\ \bibinfo {pages} {399} (\bibinfo {year}
  {1974})}\BibitemShut {NoStop}%
\bibitem [{\citenamefont {Vigolo}\ \emph {et~al.}(2014)\citenamefont {Vigolo},
  \citenamefont {Radl},\ and\ \citenamefont {Stone}}]{Vigolo2014}%
  \BibitemOpen
  \bibfield  {author} {\bibinfo {author} {\bibfnamefont {D.}~\bibnamefont
  {Vigolo}}, \bibinfo {author} {\bibfnamefont {S.}~\bibnamefont {Radl}}, \ and\
  \bibinfo {author} {\bibfnamefont {H.~A.}\ \bibnamefont {Stone}},\ }\href
  {\doibase 10.1073/pnas.1321585111} {\bibfield  {journal} {\bibinfo  {journal}
  {Proc. Natl. Acad. Sci.}\ }\textbf {\bibinfo {volume} {111}},\ \bibinfo
  {pages} {4770} (\bibinfo {year} {2014})}\BibitemShut {NoStop}%
\end{thebibliography}%

\end{document}